\documentclass[runningheads,a4paper]{llncs}

\usepackage{amssymb}
\usepackage{graphicx}

\usepackage{mathtools}
\usepackage{amsmath}
\usepackage{amssymb}
\usepackage{pgfplots}
\usepackage{pgfplotstable}

\usepackage{url}
\urldef{\email}\path{d.fanoyela@qmul.ac.uk}

\newcommand{\keywords}[1]{\par\addvspace\baselineskip
\noindent\keywordname\enspace\ignorespaces#1}

\DeclareMathOperator{\median}{median}

\begin{document}

\mainmatter  

\title{Does k Matter? \\ k-NN Hubness Analysis for Kernel Additive Modelling Vocal Separation}

\titlerunning{Does k matter?}

%
%
\author{Delia Fano Yela
\thanks{This work was funded by EPSRC grant EP/L019981/1}
\and Dan Stowell \and Mark Sandler}
\authorrunning{Fano Yela, Stowell, Sandler}

\institute{Queen Mary University of London\\
Mile End Road, London E1 4NS, UK\\
\email
}

%
%

\toctitle{Lecture Notes in Computer Science}
\tocauthor{Authors' Instructions}
\maketitle

\begin{abstract}
\emph{
Kernel Additive Modelling (KAM) is a framework for source separation aiming to explicitly model inherent properties of sound sources to help with their identification and separation. KAM separates a given source by applying robust statistics on the selection of time-frequency bins obtained through a source-specific kernel, typically the $k$-NN function. Even though the parameter $k$ appears to be key for a successful separation, little discussion on its influence or optimisation can be found in the literature. Here we propose a novel method, based on graph theory statistics, to automatically optimise $k$ in a vocal separation task. We introduce the $k$-NN hubness as an indicator to find a tailored $k$ at a low computational cost. Subsequently, we evaluate our method in comparison to the common approach to choose $k$. We further discuss the influence and importance of this parameter with illuminating results.}

\end{abstract}

\section{Introduction}

Source separation is a discipline aiming to isolate different sources from a given observable mixture. Amongst the methods for music source separation in a blind underdetermined scenario (less observable mixtures than sound sources), the major goal becomes to find inherent characteristics of the sources of interest to help with their identification and separation.

In the last decade, a number of computationally inexpensive methods explicitly modelling the target source's properties have gotten some attention \cite{FitzGerald12_MedianVocal_ISSC,RafiiPardo13_REPET_IEEE-TASLP,Fitzgerald10_HarmPercSep_DAFX,FanoYelaEFS17_HybridKamNmf_ICASSP,FanoYelaEFS17_TemporalContext_AES,FanoYelaEOS18_ShiftInvariant_ICASSP,RafiiPardo12_VoiceSepSimilarity_ISMIR}. These methods can be understood as instances of the wider kernel additive modelling (KAM) framework \cite{LiutkusFRPD14_KernelAdditive_IEEE-TSP}. The basic idea behind KAM relies on the repetitive nature of music by estimating the target source at a particular point based on points at which the source's output is somehow similar. This is typically applied to time-frequency bins in a spectrogram representation.
The function determining the target source similarity between time-frequency bins, while ignoring the entries associated with other sources, is the so-called kernel function. Consequently, if a magnitude of a bin deviates amongst the ones judged to be similar by the target source kernel, one can assume there is another overlaying source and employ order statistics to attenuate its influence.

KAM has been successfully employed for a variety of tasks in source separation, such as vocal separation, speech enhancement, percussive/harmonic separation or interference reduction \cite{LiutkusFRPD14_KernelAdditive_IEEE-TSP,RafiiP13_RepetSimSpeech_ICASSP,FanoYelaEFS17_HybridKamNmf_ICASSP}. In the case of vocal separation, a popular approach is to assume the accompaniment music to be typically more repetitive and dense compared to the vocals, considered to be sparse and varied \cite{RafiiPardo13_REPET_IEEE-TASLP}. Meaning there are more segments in the mix containing the same or similar background music than there is for vocals. The nature of these segments vary amongst methods, such as a single repeating periodic musical pattern \cite{RafiiPardo13_REPET_IEEE-TASLP}, the temporal context surrounding every time frame \cite{FanoYelaEFS17_TemporalContext_AES} or just a single time frame \cite{FitzGerald12_MedianVocal_ISSC}. In all of these cases, the background music is implicitly assumed to have a higher energy contribution than that of the vocal source. 

Amongst these methods, a popular choice for the accompaniment proximity kernel is the $k$ nearest neighbours (k-NN) function, returning the $k$ most similar frames to a given frame. The proximity measure between frames is typically based on the Euclidean distance, and therefore, two frames will be considered to be similar if they share the same centre frequency. Within the k-NN frames selection, if the vocal is indeed sparse it should appear as an outlier and can therefore be separated from the more common source through median filtering across similar bins. Since the breakdown point of the median operator is of 50\% of outliers (vocals), one could expect the choice of $k$ to be key for a successful separation. However, there is little or no guidance on how to set this parameter in the literature, nor explanation of its overall influence. 

Here we investigate the influence of the parameter $k$ in a vocal separation task and we further propose a novel method for its automatic optimisation, based on consideration of the proximity graph, which is lightweight and needs no prior training. In section 2 will introduce the KAM vocal separation baseline and discuss typical methods to choose the parameter $k$ in the K-NN proximity kernel. We will then propose a novel computationally inexpensive method for $k$ optimisation in section 3 based on graph theory statistics. In section 4 we will further analyse and discuss the impact of this parameter through an experimental evaluation and validate the proposed method in such scenario.

\section{Vocal Separation Using k Nearest Neighbours}
KAM is a framework capable of combining different approaches to source separation using different assumptions to model sound sources. From the different proximity kernel families described in  \cite{LiutkusFRPD14_KernelAdditive_IEEE-TSP}, we will focus on the models for repetitive patterns in a vocal separation task. In particular, we present a subset that can be regarded as an instance of KAM using only one iteration of the kernel backfitting procedure described in \cite{LiutkusFRPD14_KernelAdditive_IEEE-TSP}, which was also used in similar form in the REPET family of methods \cite{RafiiPardo13_REPET_IEEE-TASLP}, and later extended to account for different repetitive patterns \cite{RafiiPardo12_VoiceSepSimilarity_ISMIR,FitzGerald12_MedianVocal_ISSC}. 

These methods take advantage of the repetitive nature of music and define a distinction between a repeating background and a sparse varied foreground.
For vocal separation in popular music the background typically corresponds to the music accompaniment and the vocals can be regarded as the sparse foreground. Therefore, one can assume that the musical accompaniment contributes to most of the energy across the frequency spectrum. We follow the method and notation described in \cite{FitzGerald12_MedianVocal_ISSC} serving as the baseline method on which we will investigate the influence and optimisation of its single inherent parameter $k$.

Formally, we define the magnitude spectrogram of a musical signal as $X \in \mathbb{R}^{M \times N}$, where $M$ is the number of frequency bins and $N$ the number of time frames. For each pair of frames $(j,\ell) \in \{1,\ldots,N\}\times \{1,\ldots,N\}$ , we then compute the squared Euclidean distance between the two corresponding columns in $X$:
\[
D_{j,  \, \ell}  = \displaystyle\sum_{m \, =  \, 1}^{M} ( X_{m,  \, j} - X_{m, \, \ell} ) ^{2}.
\]
The result is a symmetric matrix $D$, which we can now sort to find the $k$ nearest neighbours to every frame by keeping track of the frame index.
Then, for every frame $j$, we create a matrix $A^j \in \mathbb{R}^{M \times K}$ containing as columns the specific subset of the $k$ most similar frames taken from $X$. 
We expect the selected $k$ closest frames to $j$ to share similar musical accompaniment and differ in terms of the vocal part. In other words, the vocal contribution in the $k$ nearest frames to $j$ can be regarded as an outlier and the musical accompaniment as the commonality between them. Consequently, the median filter is the operator of choice in \cite{FitzGerald12_MedianVocal_ISSC} to extract the common background music and separate out the vocal contribution on each frame. The estimated magnitude spectrogram $Y \in \mathbb{R}^{M \times N}$ of the musical accompaniment is:
\[
Y_{m, \, j} := \median( A^j_{m,1},\ldots,A^j_{m,K} )
\]
To extract both magnitude and vocals from the mixture, we use the soft mask $W \in [0,1]^{M\times N}$ described in\cite{FitzGerald12_MedianVocal_ISSC}. 
The complex spectrograms for the accompaniment and vocals can then be estimated by applying soft masks $W$ and $(1-W)$ respectively to the original mixture spectrogram using an element-wise multiplication.

A successful separation between background music and vocals relies largely on the vocals actually being outliers within the selection of the $k$ closest frames. We want to make sure that the $k$-NN frames have similar background music with no or different vocals. 
However, there are also frames containing matching background music \textit{and} matching vocals, which will then be very likely to be selected as near neighbours. Those frames are unhelpful for the median filtering but since the breakdown point of the median operator is of 50\% of outliers (vocals), the method is robust to the vocal repetitions up to a point. This robustness is closely related to the number of nearest neighbours we choose, i.e. the parameter $k$.  

There seems to be little or no indication on the method to find the optimal parameter $k$ in the literature \cite{LiutkusFRPD14_KernelAdditive_IEEE-TSP,FitzGerald12_MedianVocal_ISSC,RafiiPardo12_VoiceSepSimilarity_ISMIR,FanoYelaEFS17_TemporalContext_AES}. In \cite{RafiiPardo12_VoiceSepSimilarity_ISMIR} the authors introduce three other parameters to set boundaries for the choice of $k$.
However, no indication was found on how to actually fix any of those parameters, including $k$. A recent extension introducing a temporal context $R$ in the proximity kernel \cite{FanoYelaEFS17_TemporalContext_AES} performs a parameter sweep to set the new $R$ parameter to the value giving the best mean metric across a dataset.

To our knowledge, there are currently two broad approaches to setting $k$: perceptual assessment or evaluation metric optimisation. 
In the first approach one simply listens to the estimates for different $k$ values and adjusts the parameter to the best sounding setting. This is the preferred method to set $k$ when there is a reduced number of songs to be processed.
The second approach relies on a metric, typically the Signal to Distortion Ratio (SDR), comparing the estimated sound sources with the ground truth. One will set $k$ to obtain the best metric result. In practice, this means a parameter sweep for different $k$ values, for which no indication was found on how to pick. In addition, the commonly used SDR measure is known to be a proxy for perceptual quality and its  precision has been criticised \cite{CanoFB2016_EvaluationQualitySound_EUSIPCO}. However, when dealing with large datasets, perceptual assessment of the results can be very time consuming. Therefore, it is more typical to use the second approach to optimise for an overall best performance. 

A parameter sweeping approach to find the optimal $k$ value has a number of disadvantages, primarily linked to the optimisation through a performance metric. Firstly, the separation performance metrics usually require to have ground truth separate tracks available, which is not always possible in an application scenario. Further, the commonly used separation performance metrics are computationally expensive \cite{VincentGF06_PerformanceMeasurement_IEEE-TASLP}, limiting the parameter sweep to a reduced number of values in a time constraint situation. In addition, optimising $k$ using an overall performance metric does not assure the best value for all songs in the dataset. Moreover, fixing the $k$ sweep values leaves no room to inform the optimisation with the track's individual properties, such as length.

Ideally we would like to be able to automatically pick $k$ in an unsupervised way for each track separately, taking into account the nature of the song and thus finding a tailored value for $k$ assuring a successful separation.
We would also like to do this without having to perform multiple runs of source separation and discarding all but one of them.

\section{Properties of the k-NN graph}

For a given music recording, the family of KAM methods we consider depends fundamentally for its behaviour on the set of nearest neighbours selected for each of the $N$ frames.
These nearest neighbour relationships can be represented as a directed graph with frames as nodes, and each node having $k$ arcs leading outward to its nearest neighbours. Note that if frame $i$ is a neighbour of frame $j$, the reverse is not necessarily true. At extreme settings, if $k=0$ then the graph has no arcs and thus no structure, while if $k=N$ the graph is fully connected and likewise exhibits no structure. What are desirable characteristics for a k-NN graph to be used in KAM?

Unlike many problems defined on a graph, in KAM we do not wish our graph to take on simple structure such as well-separated clusters: instead, we want all frames to have connections to frames which are similar according to the current source kernel, but dissimilar in terms of the other sources.
It is not clear how these structural considerations can best be quantified numerically, though such structure would have some impact on summary statistics considered in graph theory.

Consider a set of frames containing a background musical phrase which is repeated often: we would expect these to form a densely connected component in the graph. The frames also containing sparsely-present and variable vocal energy would be expected to have arcs pointing to that densely connected component but few arcs pointing back out to them.
Therefore, the number of incoming arcs (i.e. in-degree) would be unevenly distributed across the nodes,
directly as a result of the observed signal properties which one assumes in KAM.

One way to analyse such properties in graph theory is the concept of `hubs', which are nodes with an unusually high in-degree \cite{radovanovicNI09_NNHubs_ACM}.
This has been of particular influence in social network theory as researchers studied effects such as `small world' phenomena, which can have important effects such as the speed at which news or illness spreads through a social network.
For a given graph, one can define summary statistics which reflect the general presence of hubs.
One referred to as the `hubness' is simply the skewness of the k-occurrence statistics, i.e. the skewness of the distribution of the in-degrees of nodes in the graph. 
Here, the k-occurrence of a frame corresponds to the number of times that frame is amongst the $k$ nearest neighbours, and the `hubness' is therefore the skewness of the distribution of all frames' k-occurrence. 
In a k-NN graph we assign a fixed number of arcs, and so the average in-degree is always $k$; however if the graph contains strong hubs then the skewness of the in-degree will be high.

In our vocal separation application in KAM it is clear that a graph with relatively \textit{high} hubness should typically be one which has appropriate structure.
We typically have very little \textit{a priori} guidance over what value of $k$ to choose, so it is advantageous that, for each track separately, we can iterate over a selection of possible $k$, inspect graph statistics such as hubness for the graphs thus produced, and select $k$ which produces the optimal statistics. 
Therefore, we here propose to select the $k$ producing the maximum hubness of the associated $k$-NN graph. 

However, in a situation where we vary $k$, the hubness $h$ will vary even in the null case of a randomly-constructed graph. (This can be seen in the extreme cases: for $k=0$ or $k=N$ the graph is symmetric and the hubness is $0$, whereas for other $k$ it can be nonzero.)
A standard null model can be generated by selecting $k$ neighbours for each frame purely at random. This is related to the classic Erd\H{o}s-R\'{e}nyi random graph except that it is directed rather than undirected \cite{erdos1959random}.
The distribution of k-occurrences in this null model follows a binomial distribution with parameters $N$ and $k/N$, leading to an expression for the expected hubness as:
\begin{equation}
	h_{\textrm{null}} = (1-2k/N) / \sqrt{ k (1-k/N) }
\end{equation}
We can thus define a normalised hubness statistic as the `excess' hubness, i.e. the raw observed hubness minus the hubness expected under the null model,
which should then be less biased than the raw hubness in selecting $k$.

The above null model is one of the simplest random graphs.
In practice, graphs constructed from high-dimensional similarity measures do not behave strictly in that fashion, and it is an ongoing research topic to model how k-NN graphs behave in general \cite{radovanovic2009nearest}.
In preliminary work we found that the general scaling of the hubness statistic was out of line (larger) than in the simple null model, and so our empirical normalisation is given as
\begin{equation}
	h_{\textrm{norm}} = \frac{h}{\max(h)} - \frac{h_{\textrm{null}}}{\max(h_{\textrm{null}})}
\end{equation}
where maxima are across the sweep of $k$ settings.

Using the maximum hubness as a metric to choose $k$ has numerous advantages:

\begin{enumerate}
\item It does not require any ground truth information
\item $k$ is optimised per track as a pre-processing step before the separation actually takes place
\item It is quick to compute so we can sweep through a lot of different $k$ values, so we can have a finer optimisation
\item The hubness has been demonstrated to have perceptual relevance for song similarity in music recommendation, suggesting that it reflects properties of the nearest neighbour graph that have impact on its applied use. However, it has not been used for frame selection in KAM and so that is to be explored here.
\end{enumerate}

\section{Experiments and Discussion}

To evaluate the proposed method, we quantitatively compare it against the standard parameter sweep for setting $k$ in KAM for a vocal separation task.
We chose to follow the vocal separation method described in \cite{FitzGerald12_MedianVocal_ISSC} with FFT size of 4096 and hop size of 1024 samples, as it represents a baseline instance of the larger KAM framework.

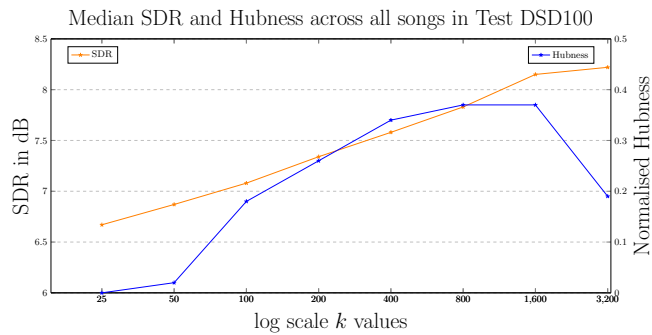
\begin{figure}[h]
\vspace{-0.3cm}
\begin{center}
\begin{tikzpicture}[scale=0.4]
\begin{semilogxaxis}[
	log ticks with fixed point,
    title={Median SDR and Hubness across all songs in Test DSD100},
    title style={font=\huge},
    xmode=log,
    scaled x ticks=real:1e3
    xlabel={ log scale $k$ values},
    ylabel={SDR in dB},
    xmin=0, xmax=3300,
    ymin=6, ymax=8.5,
    xtick={0,25,50,100,200,400, 800, 1600, 3200},
    ytick={6, 6.5,7,7.5,8,8.5},
    legend pos=north west,
    ymajorgrids=true,
    grid style=dashed,
    axis y line*=left,
    ylabel near ticks,
    label style={font=\huge},
    width = 20cm,
	height = 10cm,
     ]
 
\addplot[color=orange, mark=star, ]
    coordinates {
    (0, 6.13) (25, 6.67) (50, 6.87) (100, 7.08) (200, 7.34) (400, 7.58)  (800, 7.83) (1600, 8.15) (3200,8.22) };
    \legend{SDR} 
    \end{semilogxaxis}   
    
    \begin{semilogxaxis}[
    log ticks with fixed point,
    scaled x ticks=real:1e3
    xmin=0, xmax=3300,
    ymin=0, ymax=0.5,
    xtick={0,25,50,100,200,400, 800, 1600, 3200},
    xlabel={ log scale $k$ values },
    ytick={0, 0.1,0.2,0.3,0.4,0.5},
    legend pos=north east,
    axis y line*=right,
    ylabel={Normalised Hubness},
    ylabel near ticks,
    label style={font=\huge},
    width = 20cm,
	height = 10cm,
     ]
    
    \addplot[ color=blue, mark=star, ]
    coordinates {
    (0, 0) (25, 0) (50, 0.02) (100, 0.18) (200, 0.26) (400, 0.34)  (800, 0.37) (1600, 0.37) (3200, 0.19) 
    };
    \legend{Hubness}
    \end{semilogxaxis}
\end{tikzpicture}
\end{center}
\caption{Median SDR and hubness across all songs in Test DSD100 for different fixed $k$ values}
\label{fig:sdrhub}
\vspace{-0.4cm}
\end{figure}

\begin{figure}[h]
\centering
\includegraphics[scale=0.3]{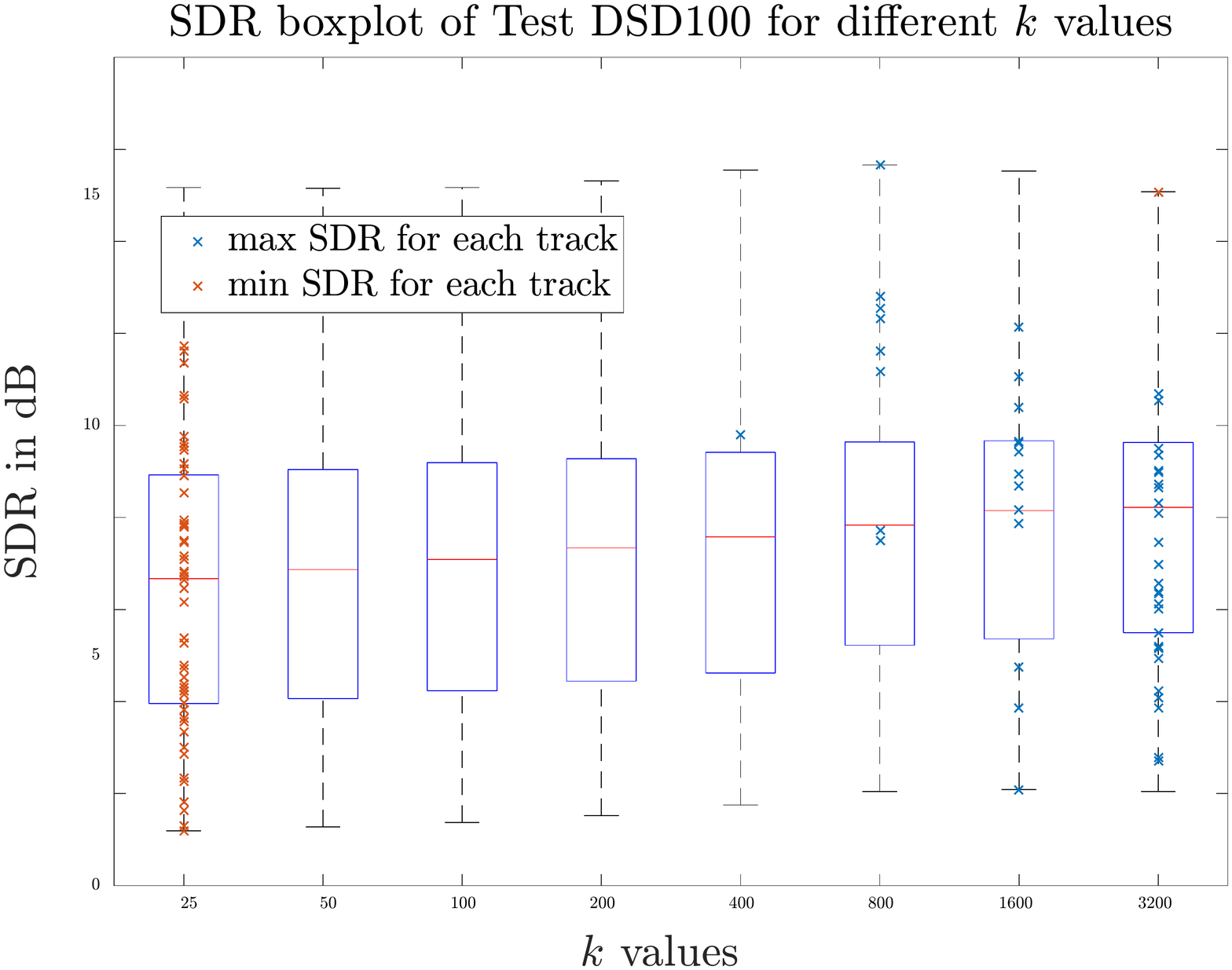}
\vspace{-0.5cm}
\caption{SDR Boxplot of every song in the Test DSD100 dataset for different $k$ values. The maximum and minimum SDR obtained for each song are marked in blue and orange respectively, showing a general trend of higher separation performance with increasing $k$ value.  }
\label{fig:sdrboxplot}
\vspace{-0.3cm}
\end{figure}

\begin{figure}[t]
\centering
\includegraphics[width=\columnwidth,height=8cm]{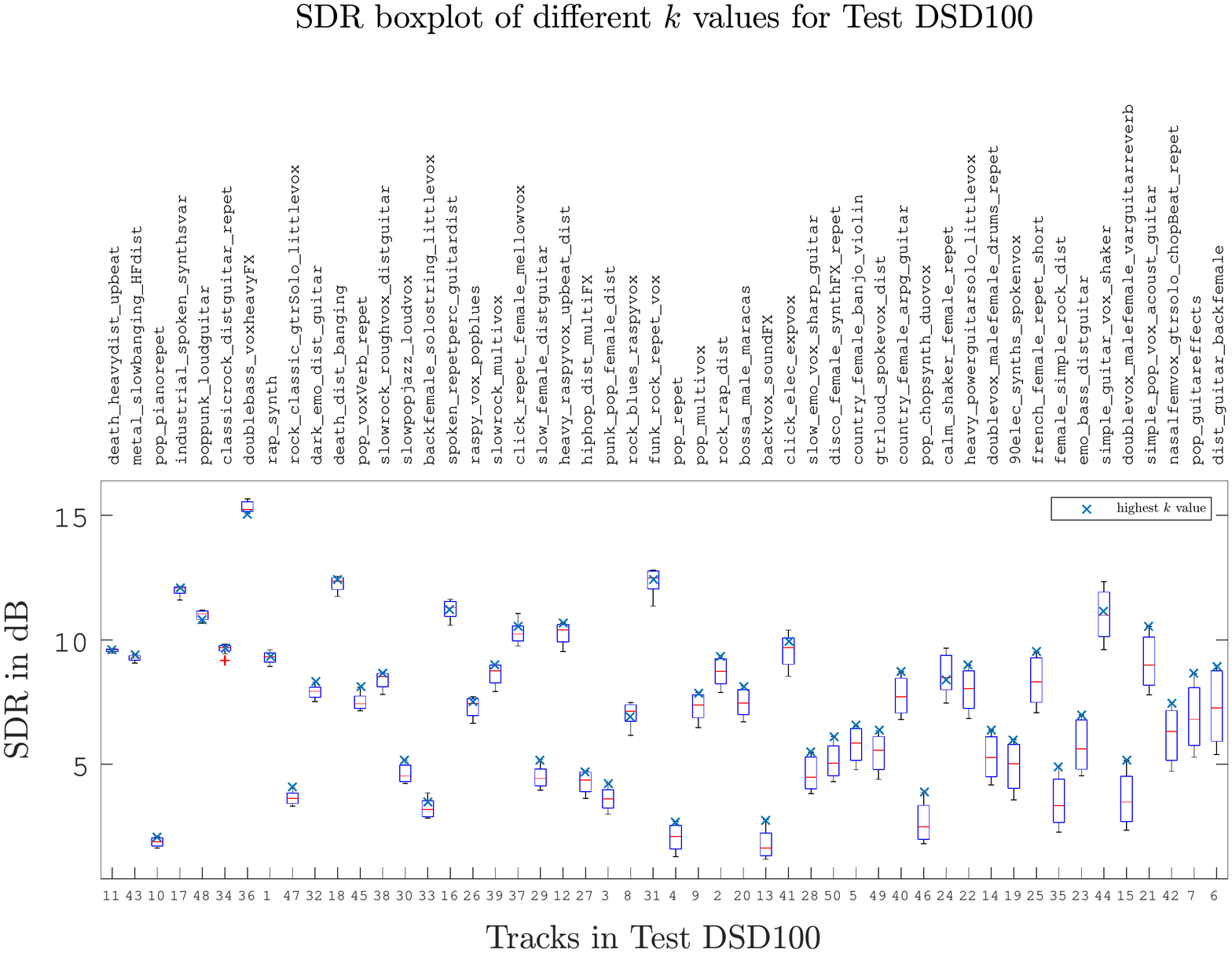}
\vspace{-0.5cm}
\caption{SDR Boxplot of different $k$ value for each song in the Test DSD100 dataset, briefly described in the top axis and sorted in ascending variance order. The SDR obtained with the maximum $k$ value of 3200 for each song is marked in blue showing the different behaviour between songs. }
\label{fig:trackboxplot}
\vspace{-0.3cm}
\end{figure}

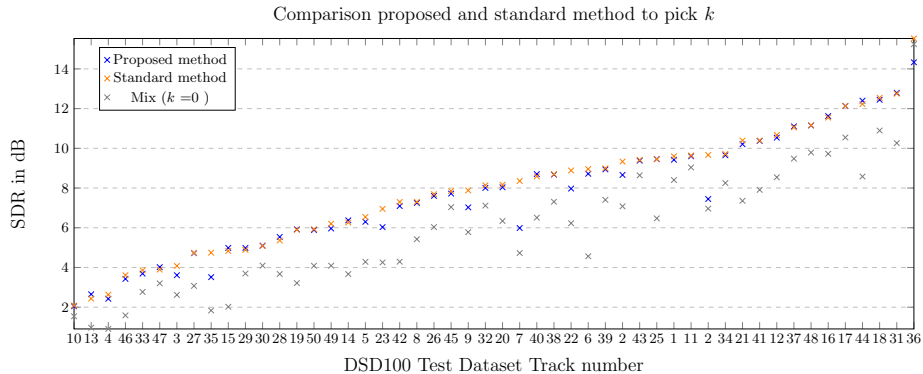
\begin{figure}[h]
\begin{tikzpicture}[scale=0.6]
\begin{axis}[
enlargelimits=false,
title={Comparison proposed and standard method to pick $k$},
title style={font=\large},
xlabel={ DSD100 Test Dataset Track number},
ylabel={SDR in dB},
label style={font=\large},
xtick={1,2,3,4,5	,6,7,	8,9,10,11,	12,13,14,15,16,17,18,19,20,21,22,23,24,25,26,	27,28,29,	30,31,32,	33,34,35,36,37,	38,39,40,41,42,	43,44,45,46,47,48,49,50},
xticklabels={10,13,4,46,33,47,3,27,35,15,29,30,28,19,50,49,14,5,23,42,8,26,45,9,32,20,7,40,38,22,6,39,2,43,25,1,11,2,34,21,41,12,37,48,16,17,44,18,31,36},              
grid style=dashed,
 ymajorgrids=true,
tick label style={font=\small},
width = 20cm,
height = 8cm,
legend pos=north west
]	
\addplot+[ color=blue, only marks, mark=x, mark size=2.5pt]
table[ ]{res_hub.dat};
\addplot+[ color=orange, only marks, mark=x, mark size=2.5pt]
table[ ]{res_sdr.dat};
\addplot+[ color=gray, only marks, mark=x, mark size=2.5pt]
table[ ]{res_min.dat};
\legend{Proposed method, Standard method, Mix ($k$ =0 )}
\end{axis}
\end{tikzpicture}
\caption{SDR values for each song in the Test DSD100 dataset sorted in ascending order, using the optimal $k$ issued from the standard and proposed method, in comparison to the SDR of the raw mixture (i.e. $k$=0).}
\label{fig:sdrtracks}
\vspace{-0.5cm}
\end{figure}

To encourage reproducibility, we use the publicly available Test Demixing Secrets Dataset (DSD100) \cite{liutkus20172016}, containing 50 full length songs of diverse genres sampled at 44.1 kHz. Since the kernel implemented relies on musical repetition, we evaluated our proposed method on full length songs to ensure as much sound material as possible for KAM's source reconstruction. However, the literature only offers some indication on $k$ values for 30 second segments. We therefore use a broad range of fix $k$ values for the traditional parameter sweep, letting $k \in \{ 0, 25, 50, 100, 200, 400, 800, 1600, 3200 \}$, and a finer percent increase sweep for the computational inexpensive proposed method taking the song length into account, letting $k \in \{ (0.001, 0.011, 0.021, 0.031, ... , 0.45) \times N \}$ where $N$ is the total number of time frames in the song. 

Following common practice in the field, we employ the Signal to Distortion Ratio (SDR) in the BSS Eval toolbox 3.0 \cite{VincentGF06_PerformanceMeasurement_IEEE-TASLP} as the quantitative indicator of the separation performance. Therefore, we would expect to observe a positive correlation between SDR and hubness for different $k$ values. 
Due to the diversity of styles in the dataset, one could also expect an improvement in the overall separation performance (and so SDR) by using a tailored $k$ for each song following the proposed method.

According to the standard method to fix $k$, one would pick the value with a higher overall SDR, here (Fig. \ref{fig:sdrhub}) is the highest $k$ of 3200 frames. Alternatively, the positive correlation between the hubness and SDR seen in Fig. \ref{fig:sdrhub} suggests the hubness to indicate the optimal $k$ value for a successful separation.

Moreover, the similarity between boxplots in Fig. \ref{fig:sdrboxplot} for different $k$ values suggests there might not be an unique $k$ that maximises the SDR of every song in the dataset.
However, the crosses indicating the $k$ value from that set for which the maximum and minimum SDR was obtained for each song partly go against this idea, as most of the separations were more successful with the highest $k$ value. This behaviour is surprising as the songs in the dataset present very distinct characteristics. One would expect most of the tracks to peak at lower $k$, since 3200 frames represents more than 30\% of the total frames for most songs, which seems to be so many frames that it should generally overpass the 50\% of outliers breaking point of the median operator. The abundance of highly repetitive songs could potentially explain how such large $k$ could be successful. 

However, the markers in Fig. \ref{fig:sdrboxplot} show differently as most of the songs obtained a higher SDR with the highest $k$ value. This behaviour comes as a surprise taking into account the dataset's disparity. Most tracks were expected to peak in SDR for lower $k$ values than 3200 frames, which seems to be so many frames that it should generally overpass the 50\% of outliers breaking point of the median operator. The abundance of highly repetitive songs could potentially explain how such large $k$ could be successful, although the literature indicates the SDR may not be a reliable metric of the actual separation performance \cite{CanoFB2016_EvaluationQualitySound_EUSIPCO}.  

Fig. \ref{fig:trackboxplot} offers a different perspective on the individual song behaviour which should shed some light on the above dilemma. 
As expected, very repetitive songs such as track 45, 4 or 50, achieve a higher SDR with highest $k$ values. However, it is also the case for unconventional pop songs such as 43 or 17, where the variance in SDR is extremely low (less than 0.05). For such cases the separation may not have been successful, but Fig. \ref{fig:sdrtracks} shows otherwise as the median SDR is above the mixture's SDR (equivalent $k=0$). Further, the overall SDR variance is surprisingly low, with a median of 1.4dB potential SDR increase by changing $k$ (maximum of 3.57dB and minimum of 0.17dB). With such a low potential SDR improvement, one might wonder if $k$ actually matters at all or again, if the SDR is failing to capture the actual separation performance. 

The majority of cases where different values of $k$ induce substantial changes in SDR correspond to popular songs with a classic pop musical set-up and repeating musical structures (Fig. \ref{fig:trackboxplot})---the ideal scenario for the implemented KAM vocal separation as described in \cite{FitzGerald12_MedianVocal_ISSC}. One could therefore infer that a track sensitive to different $k$ values (i.e. higher SDR variance), fulfills KAM requirements for a successful source separation. Track 44 presents an excellent example as it has a high SDR median and high SDR variance (2.72 dB of potential SDR improvement). However, most of the tracks in the dataset fail to present such characteristics, introducing a question regarding the flexibility and adaptability of the implemented KAM for vocal separation.

Songs which fulfill KAM ideal requirements for vocal separation (sensitive to $k$ or highly repetitive) are expected to present higher SDR values than more complex songs. 
However, Fig. \ref{fig:trackboxplot} does not present such logic, which makes one further wonder if the choice of separation performance metric is the adequate choice and so perceptual models or listening tests should be adopted for separation methods evaluation. 

Nevertheless, Fig. \ref{fig:sdrtracks} shows the proposed method can be used as substitute to the current technique for fixing $k$. Both methods present similar results in most cases and although the proposed one presents lower SDR for some songs, it seems a small trade-off for a considerable decrease in computation time (1000 times faster than the standard method).

\bibliographystyle{splncs}
\bibliography{referencesMusic,refsnew}

\end{document}